\documentclass{osa-article}

\journal{osajournal}

\articletype{Research Article}

\begin{document}

\title{Design of a passively mode-locking whispering gallery mode microlaser}

\author{Tomoki S. L. Prugger Suzuki,\authormark{1} Ayata Nakashima,\authormark{1} Keigo Nagashima,\authormark{1} Rammaru Ishida,\authormark{1} Riku Imamura,\authormark{1} Shun Fujii,\authormark{1,2} Sze Yun Set,\authormark{3} Shinji Yamashita,\authormark{3} and Takasumi Tanabe\authormark{1,*}}

\address{\authormark{1}Department of Electronics and Electrical Engineering, Faculty of Science and Technology, Keio University, Yokohama, Kanagawa, 223-8522, Japan\\
\authormark{2}Quantum Optoelectronics Research Team, RIKEN Center for Advanced Photonics, Saitama 351-0198, Japan\\
\authormark{3}Research Center for Advanced Science and Technology, The University of Tokyo, Tokyo 153-8904, Japan}

\email{\authormark{*}takasumi@elec.keio.ac.jp} 



\begin{abstract}
Ultrahigh repetition rate lasers will become vital light sources for many future technologies; however, their realization is challenging because the cavity size must be minimized. Whispering-gallery-mode (WGM) microresonators are attractive for this purpose since they allow the strong light-matter interaction usually needed to enable mode-locking. However, the optimum parameter ranges are entirely unknown since no experiments have yet been conducted. Here, we numerically investigate pulsed operation in a toroidal WGM microresonator with gain and saturable absorption (SA) to study the experimental feasibility. We show that dispersion is the key parameter for achieving passive mode-locking in this system.  Moreover, the design guideline provided in this work can apply to any small resonators with gain and SA and is not limited to a specific cavity system.
\end{abstract}

\section{Introduction}
Mode-locked ultrashort pulse lasers are vital light sources for laser processing\cite{Kerse2016}, optical communication\cite{Mollenauer:00,Ogura2001}, signal processing\cite{Cotter1523,callahan2012photonic}, LiDAR application, and remote sensing\cite{210439,DeYoung:10}. Glass-based fiber lasers are excellent platforms for these applications because of their high beam quality, robustness, and simple configuration\cite{Martinez2013}. The two key elements needed to build a mode-locked laser are a gain medium and a mode-locker, and erbium (Er) doped fiber and carbon nanotubes (CNTs), respectively, are frequently used for these purposes in a fiber laser.

For the applications mentioned above, the use of a pulsed laser with a high repetition rate will open a new avenue to faster processing, higher capacity, and more effective signal acquisition. However, realizing a repetition rate exceeding 1 GHz remains a challenge\cite{4663586} since the laser cavity length must be minimized so that the light pulses are densely packed in time. Multi-gigahertz operation has been demonstrated by using a 5-mm Fabry-P\'{e}rot Er-doped fiber cavity, where the end surface is coated with CNTs as a saturable absorber (SA)\cite{1411864,Martinez:11}, but further scaling of the repetition rate is not possible because the high pump power needed to compensate for the low gain damages the CNTs.

On the other hand, an ultrahigh-$Q$ whispering-gallery-mode (WGM) microresonator has proven to be an excellent platform that allows light to interact strongly with materials\cite{Tanabe_2019}. A WGM microresonator made of $\mathrm{SiO_2}$ on silicon has exhibited one of the highest $Q$ values of larger than $\sim10^8$.  Since it is made of $\mathrm{SiO_2}$, it is also a good host for Er doping and will eventually provide a good platform on which to demonstrate a WGM microlaser\cite{doi:10.1063/1.1873043,PhysRevA.70.033803}. Continuous-wave (CW) lasing has been demonstrated at an ultralow threshold power thanks to the high $Q$.

Moreover, dissipative soliton generation has also been demonstrated with this type of WGM resonator by applying CW pumping to one of the longitudinal modes.  Four-wave mixing in a high-$Q$ WGM microresonator generates an optical frequency comb.  If a specific condition is met, we can lock all the generated comb modes and form ultrashort pulses.  They are attractive because dissipative solitons from a WGM microresonator have an extremely high repetition rate thanks to their small cavity size.  However, such solitons sit on top of a CW background, making it difficult to achieve the pulse amplification that is often required if we are to use this light source as the seed for high-power applications.  Moreover, sophisticated wavelength sweeping and feedback are needed to generate and stabilize the soliton pulse\cite{Herr2014}, and this makes the system complex and expensive.

If we can directly obtain mode-locked pulses from a small WGM microlaser, we would achieve background-free operation without using a complex feedback system. In addition to the CW lasing operation in a silica toroid WGM microresonator, SA is also confirmed with the same device by using CNTs. So, the combination of Er-doped $\mathrm{SiO_2}$ toroid WGM microresonator with CNT as the SA is of interest and has great potential in terms of realizing an ultracompact high-repetition-rate mode-locked pulse laser.

However, it is unclear whether we can demonstrate mode-locked operation in such a device since no experiments have yet been conducted with $\mathrm{SiO_2}$ WGM microresonators for such a purpose and the optimum parameter ranges are entirely unknown. Therefore, we aim to reveal the optimum range of parameters, such as the cavity size, $Q$, Er ion concentration, dispersion, and nonlinear loss of the CNTs, needed to achieve mode-locking, and discuss the experimental feasibility.

Similar studies have already been published where researchers have tried to achieve an integrated mode-locked laser on a silicon substrate, but we would like to emphasize that we are dealing with different regimes. The technological advances made with microresonators allow us to use ultrahigh $Q$ and sophisticated Er doping in silica on silicon. We expect the high $Q$ to compensate for the limited gain of Er hence allow us to make the cavity size extremely small.

The paper is organized as follows. Section 2 describes the theoretical model and explains how we determined the parameters that we used in our numerical analysis. Some of the essential parameters taken from our experiments are explained in detail. Section 3 is the core part of this paper, where gain, nonlinear loss, and dispersions are studied as mode-locking parameters. In particular, we reveal the tradeoff between the gain and larger dispersion when we increase the cavity size. Additional numerical calculations are performed, which set design guidelines to overcome the tradeoff. Finally, we propose a design for passively mode-locking a WGM microlaser.

\section{Theoretical model and experimental verification of parameters}
\subsection{Model}
Conceptually, a WGM microlaser is a miniaturized fiber ring laser. The system that we studied is shown in Fig. 1. The resonator is made of Er-doped $\mathrm{SiO_2}$ to enable gain and is fabricated by using the sol-gel method. CNTs are deposited on the surface of the resonator as a saturable absorber. A continuous-wave pump is injected into the cavity, and the output light is coupled out by using a tapered fiber. Throughout this paper, we define the intrinsic and loaded $Q$s as follows.  The intrinsic $Q$ ($Q_{\mathrm{int}}$) is the $Q$ determined by the losses, such as the scattering and absorption of the resonator. The excess absorption losses caused by the CNTs and Er ion doping are normally excluded unless specified because they are usually considered separately in equations. The loaded $Q$ is defined as $Q_{\mathrm{load}}^{-1} = Q_{\mathrm{int}}^{-1} + Q_{\mathrm{ext}}^{-1}$, where $Q_{\mathrm{ext}} = \omega_0 \kappa^{-2}$ is the $Q$ value determined by the amplitude coupling coefficient between the resonator and the waveguide ($\kappa$) and the angular frequency of the cavity resonance ($\omega_0$).
\begin{figure}[ht!]
	\centering\includegraphics{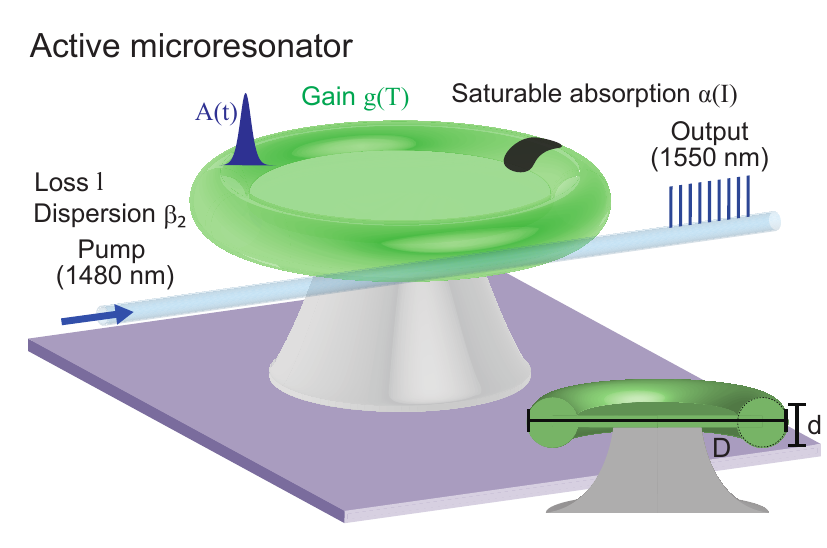}
	\caption{Er-doped silica microtoroid resonator coupled with a tapered fiber that we studied numerically. CNTs are attached to the surface of the resonator to enable saturable absorption.}
\end{figure}

We solve the model based on the nonlinear Schr\"{e}dinger equation, as given by\cite{686719,Haus:91},
\begin{equation}
	T_\mathrm{r}\frac{\partial}{\partial T}A\left( t,T\right) =\left( -iL\frac{\beta_2}{2}\frac{\partial^2}{\partial t^2} + iL\gamma\left| A\right|^2\right) A\left(t,T\right)  + \left\lbrace g_{\mathrm{T}_\mathrm{r}}\left(T\right) - \left[ l_{\mathrm{T}_\mathrm{r}} + \alpha_{\mathrm{T}_\mathrm{r}}\left( t,T\right) \right] \right\rbrace A\left( t,T\right) 
\end{equation}
where $A$, $t$, $T$ are the slowly varying field envelope in the microcavity, short time, and long time, respectively. $L$, $T_{\mathrm{r}}$, $\beta_2$, $\gamma$, $l$, $g_{\mathrm{T}_{\mathrm{r}}}$, and $\alpha_{\mathrm{T}_{\mathrm{r}}}$ are the cavity length, one roundtrip time, second-order dispersion, nonlinear coefficient, linear loss per roundtrip (determined by the loaded cavity $Q$), net-gain of Er per roundtrip, and nonlinear loss of the CNTs, respectively. Here, the nonlinear coefficient $\gamma$ was calculated using the nonlinear refractive index of silica $n_2 = 2.2\times10^{-20}~\mathrm{m^2/W}$.
\begin{equation}
	g_{\mathrm{T}_\mathrm{r}}\left(T\right) = g_0\left( \frac{1}{1+\frac{\overline{\left| A(t,T)^2\right|}}{P_{\mathrm{sat}}^g}}\right) \left(1 + \frac{1}{{\omega_\mathrm{g}}^2}\frac{\partial^2}{\partial t^2} \right) 
\end{equation}
\begin{equation}
	\alpha_{\mathrm{T}_\mathrm{r}}\left(t,T\right) = \alpha_\mathrm{ns}+\alpha_0 \frac{1}{1+\frac{\overline{\left| A(t,T)^2\right|}}{P_{\mathrm{sat}}^\alpha}}
\end{equation}
where $g_{\mathrm{0}}$, $P_{\mathrm{sat}}^g$, $\omega_{\mathrm{g}}$, $\alpha_{\mathrm{ns}}$, $\alpha_0$, and $P_{\mathrm{sat}}^{\alpha}$ are the saturated net gain coefficient per roundtrip (hereafter referred to as gain), gain saturation power, gain bandwidth, non-saturable loss per roundtrip, modulation depth per roundtrip, and saturation power, respectively.  ($\overline{|A(t,T)^2|}$) is the field intensity averaged over a short time $t$ (i.e., average power in the cavity).  We use standard split-step Fourier method\cite{agrawal2007nonlinear} to perform numerical simulations with a step size equal to the roundtrip time $T_{\mathrm{r}}$, for $1.5 \times 10^6$ roundtrips.

We must verify our numerical parameters with references and experiments before performing our numerical simulation.

\subsection{Parameters}
Here we describe the parameters of the system that we are going to use. The values are summarized in Tab. 1.
\begin{table}[htb]
	\begin{center}
		\caption{Parameter values used in simulation}
	\begin{tabular}{lllll} \hline
		Parameter & Variable & Value & Unit & Source \\ \hline
		Nonlinear coefficient & $\gamma$ & $4.2 \times 10^{-3}$ & $\mathrm{W^{-1}m^{-1}}$ & $\gamma = n_2\omega_0/cA_\mathrm{eff}$ \\
		Second order dispersion & $\beta_2$ & var. & $\mathrm{ps^2/km}$ & Calculation (Fig.~2) \\
		Quality factor & $Q$ & $10^6 - 10^8$ & - & Experiments \\
		Gain & $g_0$ & $10^{-4} - 10^{-1}$ & /roundtrip & Experiments (Fig.~3) \\
		Gain saturation power & $P_{\mathrm{sat}}^g$ & 0.145 & W & Eqs. (7)-(10) \\
		Gain bandwidth & $\omega_g / 2\pi$ & 2.5 & THz & Ref.~\cite{YARUTKINA201526} \\
		Modulation depth & $\alpha_{\mathrm{T}_{\mathrm{r}}}$ & $10^{-4} - 10^{-1}$ & /roundtrip & Experiments (Fig.~4) \\
		Non saturable loss & $\alpha_{\mathrm{ns}}$ & 0 & /roundtrip & N.A. \\
		Saturation power & $P_{\mathrm{sat}}^\alpha / A_{\mathrm{eff}}$ & 15 & $\mathrm{MW/cm^2}$ & Experiment (Fig.~4)\\ \hline
	\end{tabular}
	\end{center}
\end{table}

Among various parameters, dispersion, gain, and modulation depth are critical and need to be carefully verified since they significantly influence the mode-locking behavior.

\subsubsection{Cavity parameters: Dispersion and Q}
Dispersion is given when the material and geometry of the cavity are determined\cite{Fujii2020}. We use an $\mathrm{SiO_2}$ toroid; hence we can calculate the dispersion by determining the major diameter $D$ and minor diameter $d$ (Fig.~1). The device is usually fabricated by laser reflow, and $D/d$ is typically kept at about 10.  Hence, we assumed $d = D/10$ and obtained the dispersion, as shown in Fig.~2, and use this value for further calculation.

\begin{figure}[ht!]
	\centering\includegraphics{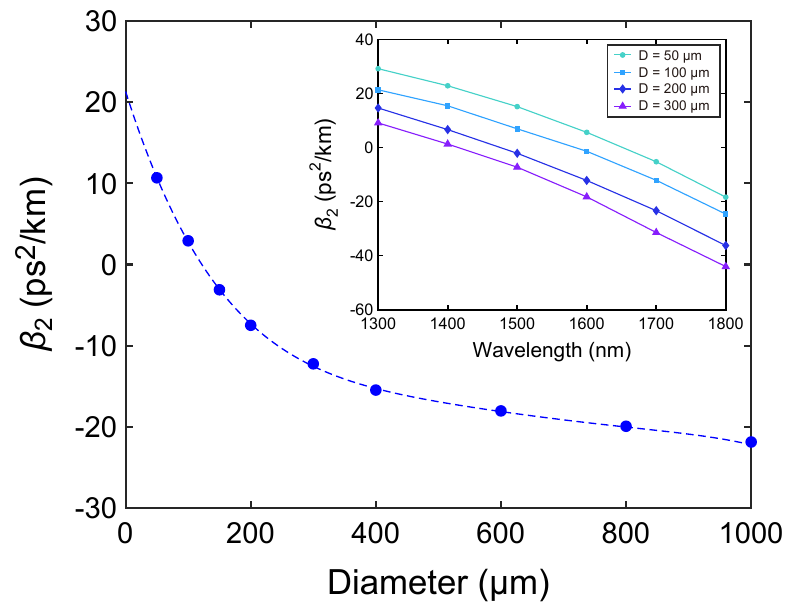}
	\caption{Second-order dispersion at 1550~nm for an $\mathrm{SiO_2}$ microtoroid with a different major diameter $D$. The minor diameter $d = D/10$. Inset is the dispersion as a function of wavelength.}
\end{figure}

As we describe in section 2.2.2, the intrinsic $Q$ is usually on the order of $10^6$ to $10^7$.  Furthermore, a cavity with an intrinsic $Q$ of $10^8$ is also possible. Hence, we study in the range from $10^6$ to $10^8$ for the intrinsic $Q$ in the $10^6$ to $10^8$ range.

\subsubsection{Er ion parameter: Gain}
We fabricated Er-doped $\mathrm{SiO_2}$ toroids by the sol-gel method\cite{Imamura:19}. We obtained a measured $Q$ of $\sim 10^7$ when no $\mathrm{Er}^{3+}$ was doped. We made two resonators, both with the same diameter of $D \sim 100$~\textmu m, but with different $\mathrm{Er}^{3+}$ concentrations of $1.7 \times 10^{18}~\mathrm{cm}^{-3}$ for device A and $0.85 \times 10^{18}~\mathrm{cm}^{-3}$ for device B. The measured loaded $Q$s (i.e., in the presence of the Er absorption effect) of these two cavities were (device A): $Q_{\mathrm{pump}} = 1.4 \times 10^6$ and $Q_{\mathrm{lase}} = 1.1 \times 10^6$, and (device B): $Q_{\mathrm{pump}} = 3.0 \times 10^5$, and $Q_{\mathrm{lase}} = 1.1 \times 10^7$, where $Q_{\mathrm{pump}}$ and $Q_{\mathrm{lase}}$ are the $Q$s of the pump and lasing modes. We pumped the cavities at $\sim$1485~nm and observed lasing at $\sim$1605~nm, where the lasing power is shown as a function of the pump power in Fig.~3(a).

When we take account of the coupled-mode theory and upper-state population of Er ions based on rate equations, the slope $\eta$ of the laser output-input power curve is theoretically given as\cite{PhysRevA.70.033803},
\begin{equation}
	\begin{split}
	\eta &= \kappa_\mathrm{s}^2\left( \frac{\nu_\mathrm{s} n_\mathrm{s} V_\mathrm{m}^\mathrm{s}}{\nu_\mathrm{p} n_\mathrm{p} V_\mathrm{m}^\mathrm{p}}\right) \left( \frac{\alpha_\mathrm{p}\left(\alpha_\mathrm{s}+g_\mathrm{s}^\ast \right) - \left(\alpha_\mathrm{p}+g_\mathrm{p}^\ast \right)\left(\alpha_\mathrm{s}+\alpha_\mathrm{s}^\mathrm{passive} \right)}{\alpha_\mathrm{s}^\mathrm{passive}\left(\alpha_\mathrm{s}+g_\mathrm{s}^\ast \right)} \right) \\
	&\cdot\left( \frac{4n_\mathrm{p}^2 \kappa_\mathrm{p}^2 \left(\alpha_\mathrm{s}+g_\mathrm{s}^\ast \right)^2}{c^2\left[\left(\alpha_\mathrm{p}+\alpha_\mathrm{p}^\mathrm{passive} \right)\left(\alpha_\mathrm{s}+g_\mathrm{s}^\ast \right)-\left(\alpha_\mathrm{p}+g_\mathrm{p}^\ast \right)\left(\alpha_\mathrm{s}+\alpha_\mathrm{s}^\mathrm{passive} \right) \right]^2}\right) 
	\end{split}
\end{equation}
where $\nu$, $n$, $V_\mathrm{m}$, $\kappa$, and $c$ are the light frequency, refractive index, mode volume, amplitude coupling coefficient between the resonator and the waveguide, and vacuum light velocity, respectively.  Scripts p and s denote pump and signal modes, respectively.  $\alpha_{\mathrm{s,p}}^{\mathrm{passive}}$ is the loss (in $\mathrm{m}^{-1}$ units) determined by the loaded $Q$ of a passive cavity without $\mathrm{Er}^{3+}$ doping for signal and pump modes. The absorption and gain (both in $\mathrm{m}^{-1}$ units) at a strong pump (i.e., $N_\mathrm{T} = N_2$ and $N_1 = 0$) are theoretically given as,
\begin{equation}
	\alpha_{\mathrm{s,p}} = \varGamma_{\mathrm{s,p}} \sigma_{\mathrm{s,p}}^{\mathrm{a}}N_\mathrm{T}
\end{equation}
\begin{equation}
	g_{\mathrm{s,p}}^\ast = \varGamma_{\mathrm{s,p}} \sigma_{\mathrm{s,p}}^{\mathrm{e}}N_\mathrm{T}
\end{equation}
where $N_\mathrm{T}$, $\sigma^\mathrm{a}$, and $\sigma^\mathrm{e}$ are the overlap factor between the optical mode with a normalized $\mathrm{Er}^{3+}$ distribution, $\mathrm{Er}^{3+}$ concentration, absorption cross-section, and emission cross-section, respectively.  For simplification we assume $\varGamma_{\mathrm{s,p}} = 1$ in our calculations.

By substituting Eqs.~(5) and (6) into Eq.~(4), we obtain graphs providing the relationship between $N_\mathrm{T}$ and $\eta$ for different $\kappa$ values, as shown in Fig.~3(b). These graphs enable us to estimate the effective $N_\mathrm{T}$ from an experimentally obtained $\eta$.

Once $N_\mathrm{T}$ is determined, we can obtain $g_{\mathrm{T}_{\mathrm{r}}}$ from Fig.~3(c). The gain can be estimated by solving the rate equation and propagation equation of an erbium-doped waveguide numerically. Hence, Fig.~3(c) is calculated as\cite{becker1999erbium}
\begin{equation}
	\frac{\partial P_{\mathrm{s,p}}\left(z,t \right)}{\partial z} = \varGamma_{\mathrm{s,p}}\left[\sigma_{\mathrm{s,p}}^{\mathrm{e}}N_2(t) - \sigma_{\mathrm{s,p}}^{\mathrm{a}}N_1(t) \right] P_{\mathrm{s,p}}\left(z,t \right)
\end{equation}
\begin{equation}
	\frac{\partial N_2(t)}{\partial t} = -\frac{N_2(t)}{\tau} - \frac{1}{A_\mathrm{eff}} \left(\frac{\partial P_{\mathrm{s}}\left(z,t \right)}{\partial z} + \frac{\partial P_{\mathrm{p}}\left(z,t \right)}{\partial z} \right) N_2(t)
\end{equation}
\begin{equation}
	N_1 + N_2 = N_\mathrm{T}
\end{equation}
where $P$, $N_1$, $N_2$, and $\tau$ are the light energy (in units of photons), ground-level carrier density, excited-level carrier density, and carrier lifetime, respectively. The gain saturates as we increase the pump power. Figure~3(c) is given by plotting the gain $g_0$ as a function of $N_\mathrm{T}$, where
\begin{equation}
	g_0 = \varGamma_{\mathrm{s}}\left[\sigma_{\mathrm{s}}^{\mathrm{e}}N_2(t_\mathrm{s}) - \sigma_{\mathrm{s}}^{\mathrm{a}}N_1(t_\mathrm{s}) \right]L 
\end{equation}
$N_{1,2}\left( t_\mathrm{s} \right) $ are the carrier densities in a steady-state calculated by using Eqs.~(7)-(9), when the pump is sufficiently strong.

\begin{figure}[htp!]
	\centering\includegraphics{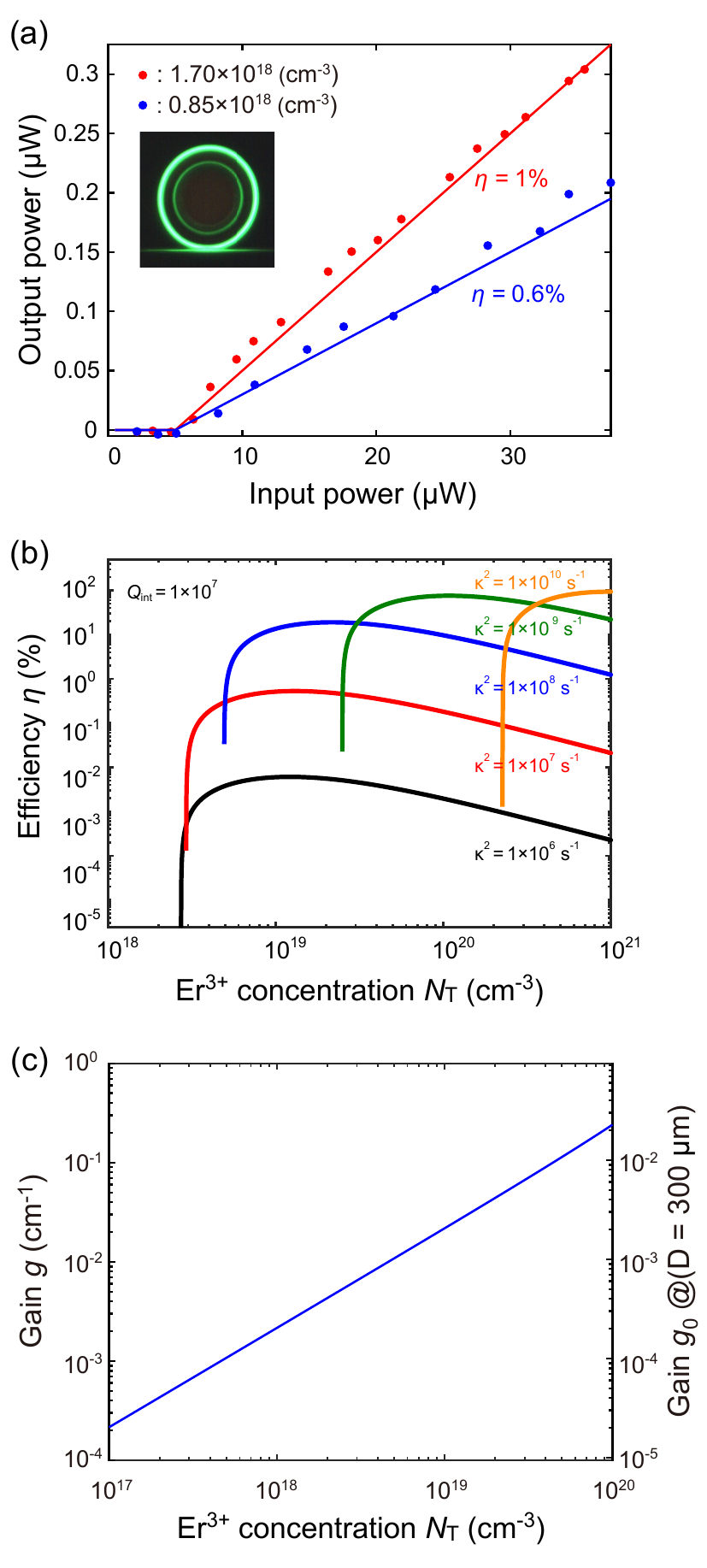}
	\caption{(a) Experimentally demonstrated laser output-input power curve for cavities with two different $\mathrm{Er}^{+3}$ concentrations $N_\mathrm{T}$. Pump power and lasing power are values in tapered fiber. We estimate $\kappa^2$ to be $10^7~\mathrm{s}^{-1}$ (under-coupling condition) from the transmittance experiment, with some uncertainness because we were not able to measure the transmittance simultaneously when performing the lasing experiment. Inset is a picture of the $\mathrm{Er}^{+3}$ doped silica toroid microresonator ($D = 100$~\textmu m) fabricated by the sol-gel method and pumped with 1480~nm laser light. (b) $\eta$ versus $N_\mathrm{T}$ for different $\kappa$ values ($\kappa=\kappa_\mathrm{p}=\kappa_\mathrm{s}$) at $\alpha_\mathrm{s,p}^\mathrm{passive} = 0.13~\mathrm{cm}^{-1}$ cavity corresponding to $Q = 10^7$ at 1550~nm. We use previously reported values of $\sigma_\mathrm{p}^\mathrm{a} = 1.5\times10^{-21}~\mathrm{cm}^2$, $\sigma_\mathrm{s}^\mathrm{a} = 2.8\times10^{-21}~\mathrm{cm}^2$, $\sigma_\mathrm{p}^\mathrm{e} = 0.8\times10^{-21}~\mathrm{cm}^2$, and $\sigma_\mathrm{s}^\mathrm{e} = 4.8\times10^{-21}~\mathrm{cm}^2$\cite{becker1999erbium}. (c) Gain $g$ (at saturating pump power) as a function of $N_\mathrm{T}$. The vertical axis on the right is the gain $g_0$ when $D = 300$~\textmu m. We use $A_\mathrm{eff} = 2.1\times10^{-7}~\mathrm{cm}^2$ corresponding to a microresonator with $d = 30$~\textmu m.}
\end{figure}

Figures~3(b) and 3(c) could be used as follows. If $N_\mathrm{T}$ is given, we can immediately obtain a theoretical $g_0$ from Fig.~3(c). On the other hand, Fig.~3(b) allows us to double-check the $N_\mathrm{T}$ value if we perform a lasing experiment, as we demonstrated in Fig.~3(a). Figure~3(a) has a slope efficiency of $\eta = 1\%$ at an experimental $\mathrm{Er}^{3+}$ doping concentration of $N_\mathrm{T} = 1.7 \times 10^{18} ~ \mathrm{cm}^{-3}$. This corresponds to an effective $N_\mathrm{T}$ of ~$6.3 \times 10^{18} ~ \mathrm{cm}^{-3}$ according to Fig.~3(b) and an experimentally estimated gain of $g_0 = 2.1 \times 10^{-3} ~ \mathrm{cm}^{-1}$ from Fig.~3(c).  On the other hand, the theoretical curve in Fig.~3(b) suggests that we should obtain a gain of $g_0 = 5.6 \times 10^{-4} ~ \mathrm{cm}^{-1}$ at an $N_\mathrm{T} = 1.7 \times 10^{18} ~ \mathrm{cm}^{-3}$ concentration. Although we do not know exactly why we obtained a slightly higher effective $N_\mathrm{T}$, we think the values are in reasonably good agreement and within the experimental error.  Taking some experimental uncertainty into consideration, we decided to investigate the $g_0 = 10^{-4}$ to $10^{-1}$ range.

Equations~(7)-(10) also allow us to calculate $g_0$ versus $P_\mathrm{s}$, at different $P_\mathrm{p}$ values, from which we can obtain $P_\mathrm{sat}\sim0.184$~W when we assume a 500~mW pump. We use this value for the calculation.

\subsubsection{SA parameters: Saturation power and modulation depth}
Saturable absorption is achieved by using CNTs. Here we explain how we verified the SA operation and obtained the SA parameters that we needed for our calculation.  The CNTs are dispersed in polydimethylsiloxane (PDMS) at a concentration of 0.24~mg/ml.  A PDMS/CNT droplet is transferred from a thin fiber to the $\mathrm{SiO_2}$ toroid, as shown in the inset of Fig.~3(b). Once the PDMS has been cured by applying heat, the absorption coefficient is measured by a transmittance measurement. The measurement method is described in further detail elsewhere\cite{Ishida:19}. The result is shown in Fig.~3(b). The fitting curve gives us a saturation power of 15~MW/$\mathrm{cm}^2$ and a modulation depth $\alpha_0$ of $3.2\times 10^{-3}$.  These values are also consistent with experiments reported by other groups\cite{doi:10.1063/1.4996918}. Therefore, we used $P_{\mathrm{sat}}^{\alpha}/A_{\mathrm{eff}} = 15$~MW/$\mathrm{cm}^2$ and studied a modulation depth range of $10^{-4}$ to $10^{-1}$.

\begin{figure}[ht!]
	\centering\includegraphics{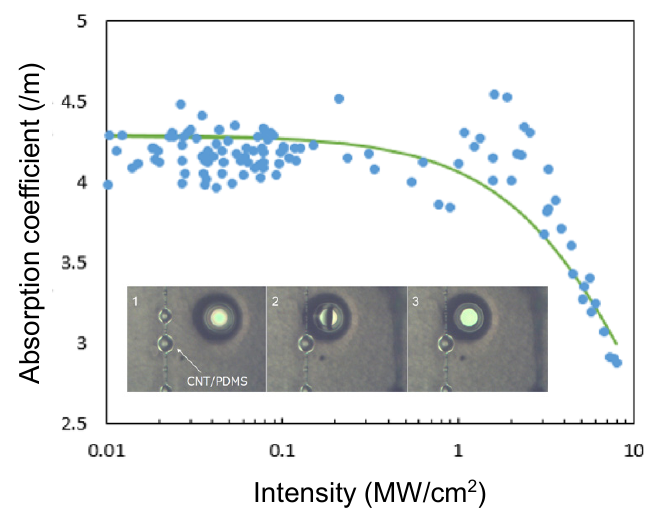}
	\caption{Absorption coefficient measurement. CNTs are dispersed in PDMS, and a droplet is transferred to the fabricated $\mathrm{SiO_2}$ toroid, as shown in the inset figure.}
\end{figure}

\section{Numerical calculation results and discussions}
This numerical study aims to reveal the optimum parameters needed to achieve mode-locking operation in a tiny Er-doped WGM microresonator. When we monitored the waveform under different conditions, we found four different states, as shown in Fig.~5.  These four states are chaotic pulses (CP), multiple pulses (MP), a stable mode-locking regime (ML), and a continuous wave (CW).  The regime that we are interested in is the ML regime, where a single pulse circulates in the WGM resonator.

We investigated the way in which the gain, loss, and dispersion of the cavity play a role in achieving stable mode-locking.

\begin{figure}[ht!]
	\centering\includegraphics{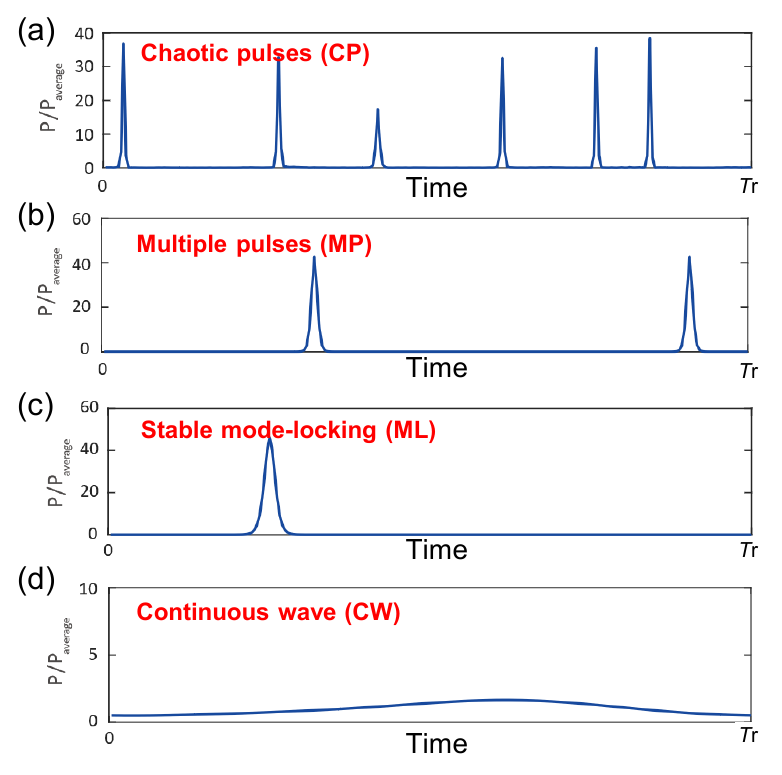}
	\caption{Operation regimes were identified based on our three criteria: chaotic pulses (CP), multiples pulses (MP), stable mode-locking (ML), and continuous wave (CW).}
\end{figure}

\subsection{Q factor dependence}
In this section, the relationship between gain per roundtrip $g_0$ and nonlinear loss (modulation depth $\alpha_0$) is investigated for an $\mathrm{SiO_2}$ toroid microresonator where $D = 300$~\textmu m. Figure~5 shows the simulation results for intrinsic $Q = 10^7$ and $10^8$.  The red line indicates the estimation of $g_0$ when $N_\mathrm{T} = 6.3 \times 10^{18} ~ \mathrm{cm}^{-3}$, which we obtained from our lasing experiment (Fig.~3).

The regimes for the four different states (CP, MP, ML, and CW) are indicated in the figure.  When the gain is low, the loss is larger than the gain, and the device will not lase.  When the gain is greater than the loss, the device starts to lase.  At a higher gain, along with an adequate modulation depth, the system exhibits ML operation. When we increase the gain even further, the pulse splits into multiple components and exhibits an MP state. Finally, the system is in an unstable regime at CP state when the gain is too high.

Figure~6(a) shows that the cavity exhibits CW lasing when $Q$ is $10^7$ at $g_0 = 10^{-3}$, which explains our CW lasing demonstration in Fig.~3(a).  However, it also shows that it is very challenging to achieve the ML state due to insufficient gain if we use this cavity. This situation will change if we use a cavity with a higher $Q$ of $10^8$, with which we should be able to reach the ML regime. Although a $Q$ of $10^8$ is experimentally feasible, it is not easy to realize.

Therefore, we will look for another way of achieving ML, such as increasing the WGM microcavity laser diameter or designing the cavity dispersion. By increasing the cavity diameter, we expect to obtain increased gain per roundtrip.  On the other hand, we expect that the mode-locking threshold will decrease if we can make a cavity that has a smaller dispersion.  Both approaches will relax the cavity loss conditions for mode-locking.

\begin{figure}[ht!]
	\centering\includegraphics{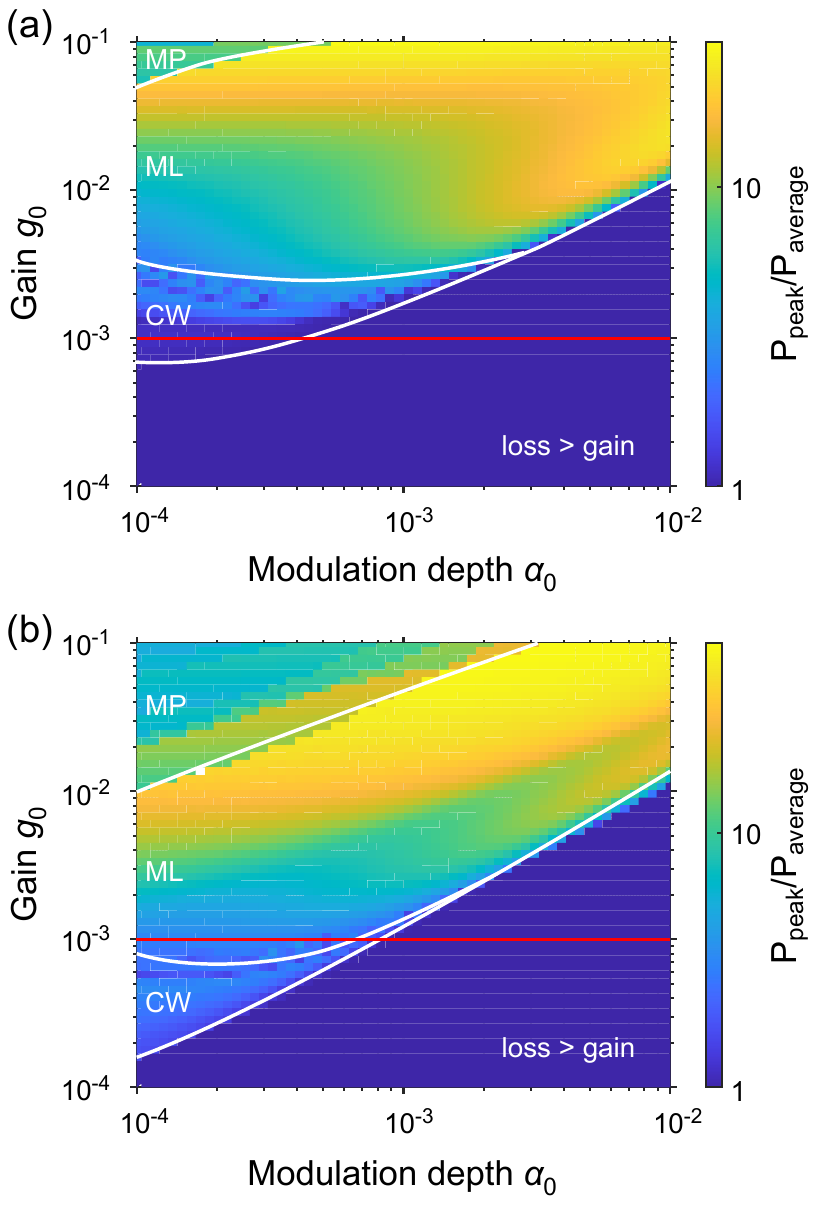}
	\caption{Mode-locking investigation of a $D = 300$~\textmu m toroidal microresonator with gain and nonlinear loss as parameters. (a) shows results for $Q = 10^7$ and (b) for $Q = 10^8$. The red line shows the position of the theoretical gain at $N_\mathrm{T} = 6.3\times10^{18}~\mathrm{cm}^{-3}.$}
\end{figure}

\subsection{Size and dispersion dependence}
As discussed previously, a sufficiently large $g_0$ is crucial for achieving ML. An alternative way of increasing $g_0$ is to increase the size $D$ of the microresonator. This approach appears to be straightforward, but the presence of dispersion makes the optimization more complex\cite{Fujii2020}. Figure~2 shows that the dispersion is at its minimum at $D \sim 150$~\textmu m, but becomes anomalous when we increase the cavity size. A smaller dispersion is usually advantageous for easy mode-locking because of the stronger self-phase modulation that occurs at a lower intracavity power. The presence of self-phase modulation is normally needed to achieve mode-locking. On the other hand, we prefer a larger cavity size because of the larger gain.  Therefore, it appears that there is a tradeoff between gain and dispersion when we change the cavity size, and there may be an optimum point.

First, we investigated only the dispersion effect. Namely, we assumed that we can fix the diameter of the toroid microresonator at 300~\textmu m but can change the dispersion. Figure~7(a) shows the numerical results we obtained when we fixed the modulation depth $\alpha_0$ and $Q$ at $5\times10^{-4}$ and $10^7$, respectively. The red line again shows an experimental estimation of the $g_0$. As expected, ML is possible even with a small $g_0$ when the dispersion is close to zero. Unfortunately, in this case, however, the dispersion of a 300~\textmu m diameter toroid is about $-20$~$\mathrm{ps}^2$/km, which is outside the ML regime.

Second, we performed further numerical investigations at different cavity diameters $D$ for a cavity where $Q$ was fixed at $10^7$.  The dispersions for a cavity with different $D$s are taken from Fig.~2.  Increasing the diameter will increase the gain per roundtrip but result in a larger dispersion. On the other hand, the decreasing diameter will reduce the gain per roundtrip but achieve a weaker anomalous dispersion. Since these two approaches appear to have a tradeoff relationship, we confirmed which design approach was more suitable for passively mode-locking a WGM micro-laser. The result is shown in Fig.~7(b).

\begin{figure}[ht!]
	\centering\includegraphics{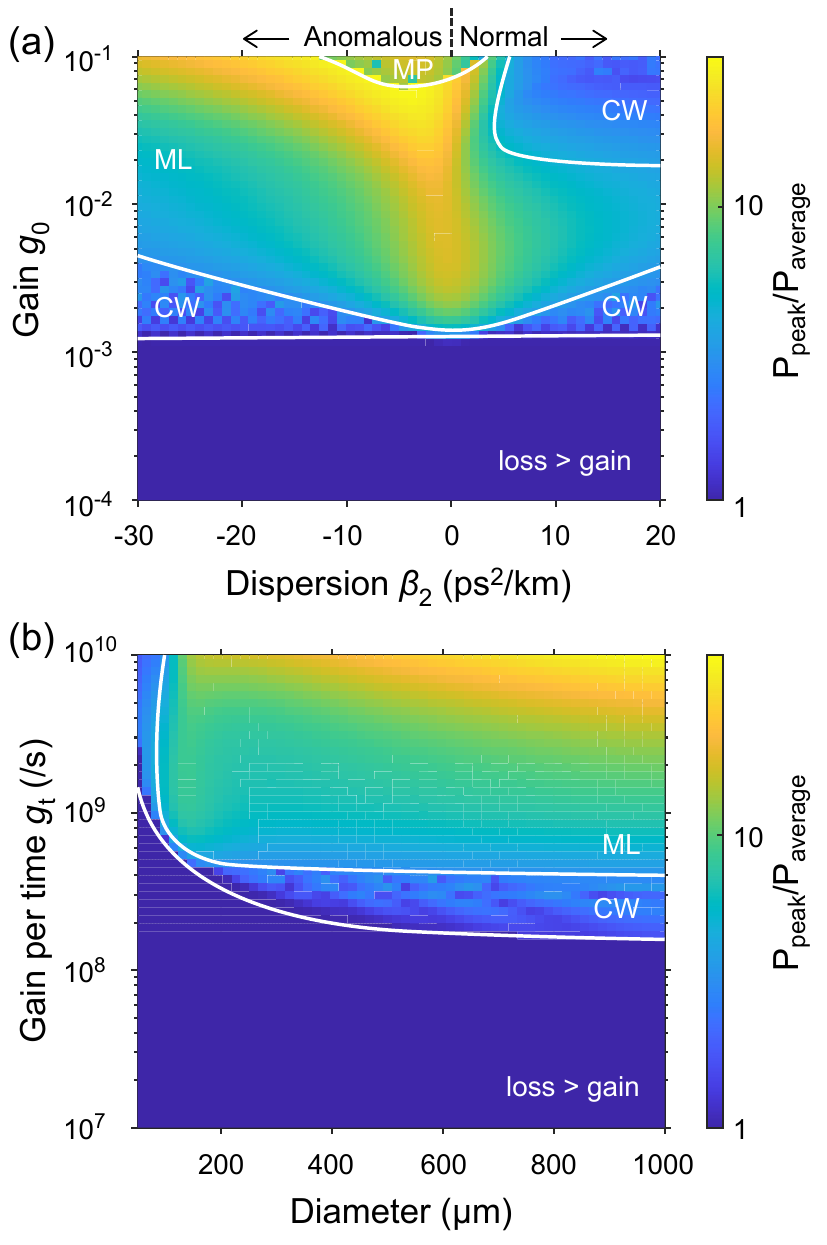}
	\caption{(a) Modelocking investigation of a 300~\textmu m diameter toroidal microresonator for $Q = 10^7$, with gain per roundtrip ($g_{\mathrm{T}_\mathrm{r}}$) and dispersion ($\beta_2$) as parameters. (b) Modelocking investigation for toroid microcavities ($Q = 10^7$) with different diameters $D$ while $D/d = 10$ is maintained. Note that the vertical axis is now gain per second ($g_\mathrm{t}$), where $g_\mathrm{t} = g_0/T_\mathrm{r}$. The value is at $g_\mathrm{t} = 2.2\times10^8~\mathrm{s}^{-1}$ when $g_0 = 10^{-3}$ and $D = 300$~\textmu m ($T_\mathrm{r} = 4.5$~ps).}
\end{figure}

Figure~7(b) shows interesting behavior. As expected, the CW lasing threshold decreases as we increase the cavity diameter since the gain increases. However, the ML threshold is not sensitive to the diameter, and it is almost constant when $D > 300$~\textmu m. This is encouraging because it is telling us that ML is possible even with a small cavity, which is usually more challenging.  Taking into consideration that a larger diameter WGM toroid resonator is often more difficult to fabricate\cite{Ma:17}, we concluded that the target diameter of the cavity is $D = 300$~\textmu m, but with a slightly higher intrinsic $Q$ and higher doping concentration.

Moreover, our results also suggested another approach, namely dispersion control. If we can shift the null dispersion point at a large $D$, we should be able to reduce the ML threshold since we can use a larger cavity.  Throughout this study, we kept $D/d = 10$, but we might change this ratio (a larger $D$ with a smaller $d$) by adjusting our laser reflow condition.

Finally, we would like to emphasize that this design guideline is not especially for this specific system but could be applied to any microcavity system with gain and saturable absorption.

\section{Conclusion}
In this paper, we numerically investigated the passive mode-locking of a toroidal WGM microlaser. The small cavity size means that ultrahigh repetition rates can be achieved; on the other hand, the gain per roundtrip is minimized, and therefore ultra-high $Q$ microresonators are necessary. Moreover, dispersion plays an essential role in mode-locking: a weak anomalous dispersion promotes ML operation with a limited gain. 

In addition, the tradeoff relationship between gain and dispersion was highlighted due to their dependence on microresonator diameter. As a result, a 300~\textmu m diameter WGM microlaser with a $Q$ slightly higher than $10^7$ proves to be a promising platform for a microlaser with an ultra-high repetition rate exceeding 100~GHz.

Finally, this design guideline can be applied to any microcavity system that has gain and saturable absorption.

\begin{backmatter}
\bmsection{Acknowledgments}
This work was supported by JSPS KAKENHI (JP18K19036, JP19H00873), Amada Foundation, and MEXT Q-LEAP.

\bmsection{Disclosures}
\noindent The authors declare no conflicts of interest.
\end{backmatter}

\bibliography{Tomoki_paper}






\end{document}